\documentclass{elsart}

\usepackage{natbib}

 \usepackage{graphicx}

\usepackage{amssymb}

\newcommand{\Lsun} {L$_\odot$}
\newcommand{\Msun} {M$_\odot$}

\begin{document}

\begin{frontmatter}



\title{Galaxies in the infrared}


\author{Ralf Siebenmorgen }

\address{European Southern Observatory, Karl-Schwarzschildstr.2,
D-85748 Garching b. M\"unchen}

\begin{abstract}

The mid infrared spectra of the starbursts show the 9.7$\mu m$
silicate absorption feature and strong mid infrared emission bands
centered at 6.2, 7.7, and 11.3 $\mu$m. Illustrative models of the {\it
active} galaxies are presented: As the starbursts are most likely
confined to the central region of the galaxy the radiative transfer in
the nucleus is solved under the assumption of spherical symmetry.  The
distribution of stars and dust are adjusted until the complete
infrared spectrum of the galaxies are modeled. The dust is described
as a mixture of large grains, very small grains and PAH, which are
undergoing temperature fluctuations. Although the galactic nuclei are
deeply hidden in the dust its global structure can be estimated by the
simple calculations presented.

\noindent
ISOPHOT samples of eleven {\it active} galaxies and seven {\it
inactive} spirals are presented. The far infrared and submillimeter
spectrum of the {\it active} galaxies can be described by a single
modified black--body at a color temperature of $31.5 \pm 2.8$\,K.
This leads to a ratio of infrared luminosity to gas mass, $L_{\rm
IR}/M_{\rm gas}$, of $\sim 90$\,L$_\odot$/\,M$_\odot$.  In contrast,
the spectral energy distributions of {\it inactive} spirals require,
apart from warm dust of $31.8 \pm 2.8$\,K, an additional very cold
component of at most $12.9\pm 1.7$\,K. This implies a $L_{\rm
IR}/M_{\rm gas}$ ratio of $\sim 3$\,L$_\odot$/\,M$_\odot$ \/ for the
{\it inactive} spirals, a factor $\sim 30$ lower than for the {\it
active} galaxies. The detection of such cold dust can be predicted by
radiative transfer models.

\end{abstract}

\begin{keyword}
Dust \sep radiative transfer \sep active galaxies \sep spirals

\end{keyword}

\end{frontmatter}

\section{Introduction}

Star formation in external galaxies are intimately linked to questions
concerning galaxy formation and evolution. Of particular interest are
galaxies with star bursts. Because formation is taken place in
interstellar clouds, the UV and optical stellar light is re-processed
into infrared photons and it is necessary to observe at
infrared/millimeter wavelengths and interpret the observations by dust
and radiative transfer models.

\begin{figure}
\centering

\includegraphics[width=1.0\textwidth]{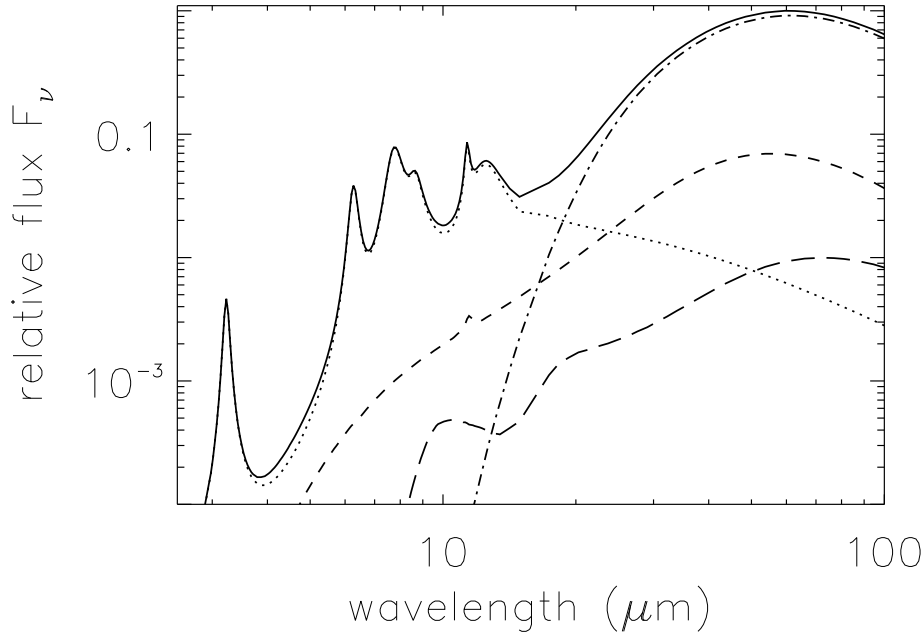}

\caption[]{Typical contribution of the individual dust populations
(Siebenmorgen \& Kr\"ugel 1992) to the total IR spectrum (full
line). The emission of large grains is shown by the dash-dotted line,
very small graphite by the dashed line, very small silicates by the
long dashed line, and PAHs by the dotted line. Figure adopted from
Siebenmorgen et al. (1998). \\}

\label{ring_sed.ps}
\end{figure}

\section{Dust model}

A two component model of the dust in the interstellar medium
(Siebenmorgen \& Kr\"ugel, 1992) has found some success in the last
years.  This model computes the emission per unit mass of dust heated
by radiation of known intensity and spectral distribution. The dust
consists of a mixture of large grains, very small grains and organic
molecules.  We assume here that the molecules are polycyclic aromatic
hydrocarbons (PAH), though other band carriers have been suggested
(e.g.  Sakata et al. 1984, Duley 1988, Papoular et al. 1989). To
consider somehow the shape of the bands a Lorentzian profile with some
damping constant was suggested by Siebenmorgen et al. (1998). Such a
simple oscillator model can account for observations taken at moderate
spectral resolution. One typical example demonstrating the importance
of PAH in the mid infrared as well as of large grains but in the far
IR is shown in Fig.\ref{ring_sed.ps}.


\begin{figure}
\centering
\includegraphics[width=1.3 \textwidth]{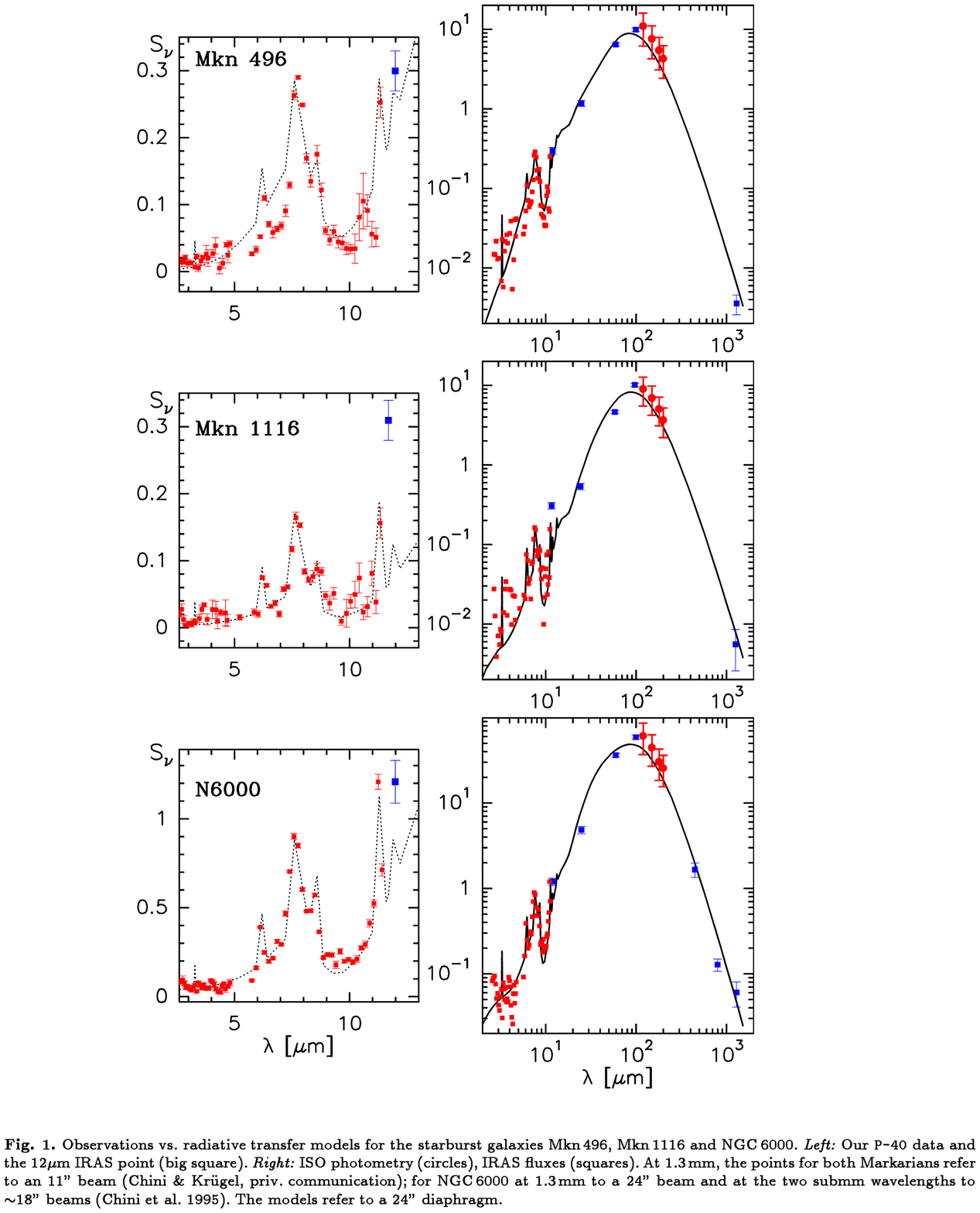}

\caption [] {}

\label{starburst.ps}
\end{figure}

\section{Radiative transfer model}

The galactic nuclei are deeply hidden by dust so that it necessary to
carry out radiative transfer calculations in some form of
approximation. We applied the numerical code by Kr\"ugel \&
Siebenmorgen (1994) which considers in a consistent computation also a
distribution of stars in the galactic nuclei. Because some of the
massive stars are so luminous each immediate environment of an OB star
presents a hot spot where the dust temperature has a local peak. At
those hot spots the heating of the dust is dominated by the stars
while outside the hot spots the dust is heated by the interstellar
radiation field (ISRF). In Fig.\ref{starburst.ps} we present some of
our ISO data together with submillimeter observations. The data are
fit quite well with the dust and radiative transfer model described
above. In particular note the strong silicate absorption best seen in
the log-log diagrams. The silicate absorption feature is quite
difficult to be recognized from the PHT-S data alone.

For all galaxies the model parameters give some tentative description
of the nucleus and its structure (Siebenmorgen et al. 1999a).
However, one should also not forget about its limitations:
Computations are done in a broken three dimensional but still
spherical grid. The real configuration of the nucleus is evidently not
spherical so that the structure of the model becomes oversimplified on
scales smaller than 400 pc.

\begin{figure}
\centering
\includegraphics[angle=90,width=1.0\textwidth]{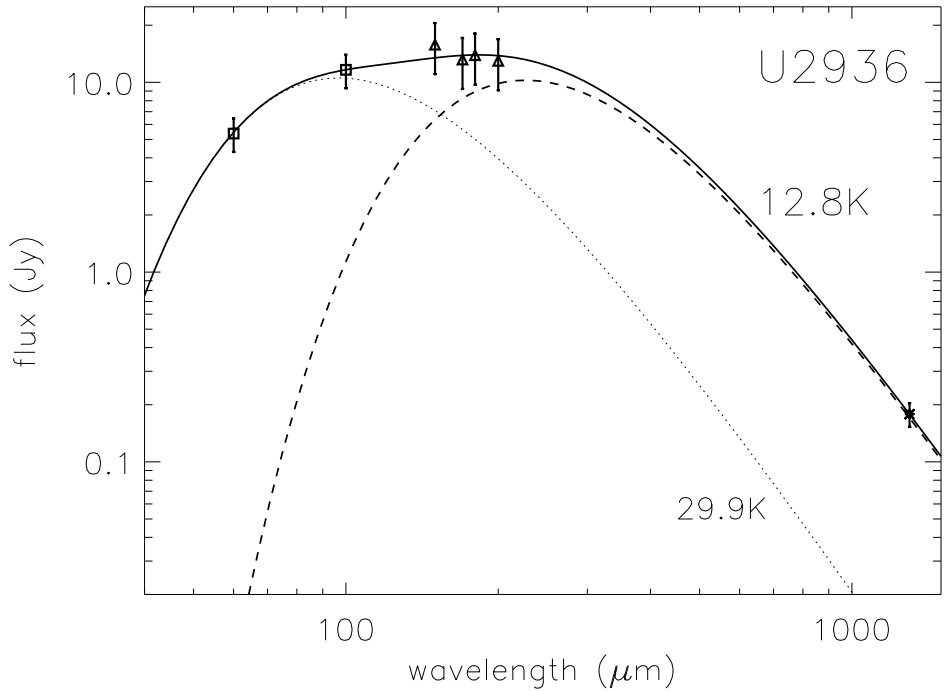}
\caption[]{ The spectral energy distribution of UGC2936 between 40 and
1500\,$\mu$m.  ISOPHOT ($\triangle$) and IRAS data are color
corrected.  The ISOPHOT error bars reflect the absolute photometric
calibration uncertainty. Photometry at 1.3\,mm by Chini et
al.~(1995). We show the best fit (solid curve) is separated into a
29.9K cold (dotted) and a 12.8K very cold dust component
(dashed). Figure adopted from Siebenmorgen (1999b).\\}
\label{U2936_sed.ps}
\end{figure}
\begin{figure}
\centering
\includegraphics[width=1.0\textwidth]{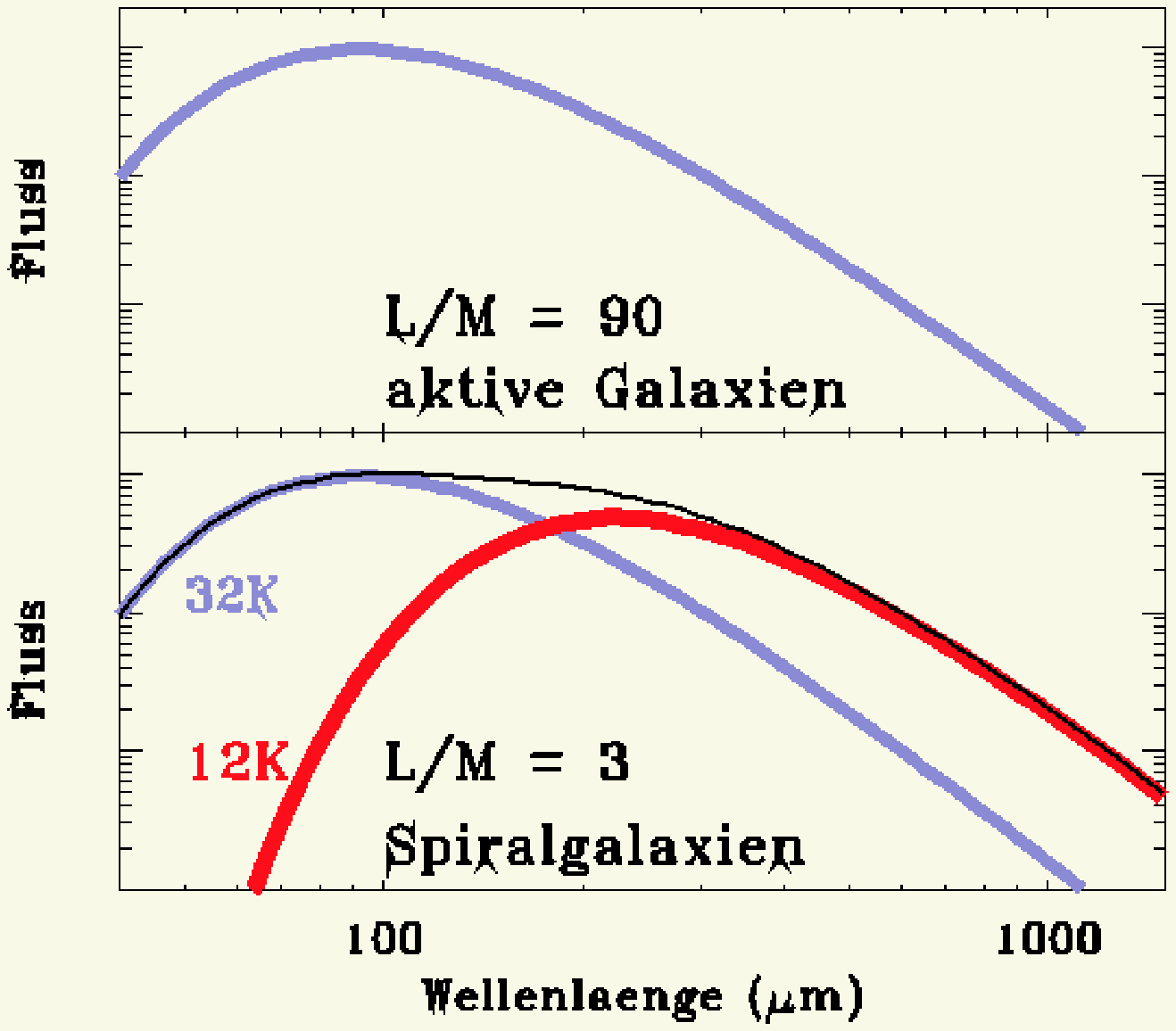}
\caption[]{Generic spectra calculated as a simple average over our ISO
sample of Markarian (top) and spirals (bottom).  The FIR/mm emission
of the Markarian galaxies can be fit by a single modified black
body. However, for all spirals an additional very cold dust component
is needed to fit the observations. The derived $L_{\rm IR} / M_{\rm
gas}$ ratio can be used as a good discriminator of the activity type
of a galaxy. \\}
\label{kalterStaub_sum.ps}
\end{figure}

\section{FIR/mm continuum emission of galaxies}

The far IR emission of the starburst galaxies presented in
Fig.\ref{starburst.ps} are well fit by a single modified black body at
about 32K. Siebenmorgen et al. (1999b) have studied the far IR/mm
emission of another sample of 8 galaxies extracted from the Markarian
Catalog (Gardner 1995). All of them show a similar FIR/mm spectral
energy distribution. The spectral energy distribution of the inactive
galaxy UGC2936 is shown Fig.\ref{U2936_sed.ps}. Our best fit which is
constrained by ISO, IRAS and IRAM data reveals the presence of second
component. We need to employ a 29.9K cold and a 12.8K very cold dust
component.

Our multi-filter far IR photometric study of 7 spirals shows that at
least two dust components are needed to account for the
spectra. Beside a cold dust component of about 32K an additional
component of very cold dust of about 12K is present.  A simple average
over all Markarian and another average over all inactive galaxies of
our ISO sample give us generic spectra for both types of
galaxies. They are shown in Fig.\ref{kalterStaub_sum.ps}. Siebenmorgen
et al. (1999b) therefore confirm the original claim by Kr\"ugel et
al. (1998) of the existence of such an additional component of very
cold dust in spiral galaxies. Determining the gas mass from 1.3mm dust
continuum maps that cover the optical extent of the inactive spirals
we find a ratio of total infrared luminosity to gas mass $L_{\rm IR} /
M_{\rm gas} \sim 3 $ \Lsun / \Msun, which is a factor 30 lower than
found for the Markarian galaxies. This demonstrates that the $L_{\rm
IR} / M_{\rm gas}$ ratio can be used as a discriminator of the
activity type of a galaxy.

\noindent
{\bf Question: {\it P. Barthel}} \\
\noindent
Could it be that cold dust is present in the MKN galaxies but warm
dust is just swamping it ?  \\
\noindent
{\bf Answer:} \\
\noindent
Both galaxy samples are at about similar distance scale and have been
observed with the same instrument sensitivities. The (very) cold dust
can be best studied in the submillimeter. There is no or only marginal
evidence for source extension at 1.3mm for the Mkn sample whereas the
spirals have similar extension as the optical diameter. Therefore the
observations do not give evidence of a very cold dust component for
the active sample but for the spirals. \\

\clearpage

\end{document}